# Trends in scientific research in *Online Information Review*. Part 1. Production, impact and research collaboration

Rafael Aleixandre-Benavent[12*]; Carolina Navarro-Molina[2]; Remedios Aguilar-Moya[3]; David Melero-Fuentes[4]; Juan-Carlos Valderrama-Zurián[2]

1. Instituto de Gestión de la Información y del Conocimiento (INGENIO) (CSIC-UPV), Spain.
2. Unidad de Información e Investigación Social y Sanitaria (CSIC-UV), Spain.
3. Departamento de Ciencias de la Educación, Universidad Católica de Valencia "San Vicente Mártir", Spain.
4. Instituto de Documentación y Tecnologías de la Información, Universidad Católica de Valencia "San Vicente Mártir", Spain.

**\*Correspondence:**
rafael.aleixandre@uv.es
Plaza Cisneros, 4
46003-Valencia, Spain

## Abstract

The study, based on the Web of Science, analyses 758 articles published from 2000 to 2014.

Our analysis includes the publications' output, authorship, institutional and country patterns of production, citations and collaboration. A Social Network Analysis was conducted to identify primary groups of researchers and institutions and the collaboration between countries.

The study reveals that 1097 authors and 453 Institutions have contributed to the journal. The collaboration index has increased progressively, and the average degree of collaboration during the study period was 1.98. The majority of the papers were contributed by professionals affiliated with a university. Highly cited papers address online and digital environments, e-learning systems, mobile services, web 2.0 and citation analyses.

This work is a bibliometric analysis of a leading journal in library and information science, Online Information Review.

## 1. Background

*Online Information Review* is an international journal devoted to research in the field of online information in academic, government, corporate, scientific and commercial contexts. The journal seeks to provide a forum in which to share the experiences of various specialists related to information science, including information technology, information management, knowledge management and related social sciences, comprising metrics for research evaluation. *Online Information Review* focuses on issues related to online systems, services and resources and their use. New trends in information research, such as alternative metrics for measuring research impact (altmetrics), data mining and text mining, human-computer interaction, mobile applications and solutions in online environments, and usability and user interfaces for information environments, are also included.

Papers published in scientific journals are one of the measurable outcomes of research activity and may be analysed using qualitative and quantitative methods. The qualitative method of "peer review" is a requirement imposed by editors for publishing in prestigious scientific journals. The quantitative methods are based on measures derived from statistical analysis of published scientific literature (White and McCain, 1989). These measures reflect the scientific activity of academics and research institutions by determining the characteristics of the published papers and the collaborative relations among authors, institutions and countries. Moreover, because authors confer recognition on colleagues' publications by citing those studies, citation counts reflect the effect of published papers on subsequent publications and their authors (Aleixandre-Benavent *et al*., 2007).

Scientific collaboration facilitates the flow of information among researchers and also allows for cost-sharing and improved efficiency in research (Kretschmer, 2004; Newman, 2004). Social network analysis allows the graphic representation of interpersonal and inter-institutional collaborations by quantifying the number of members who are included in a network, the intensity of their relationships and which members are the most relevant (González-Alcaide *et al*., 2008a; González-Alcaide *et al*., 2008b; Newman, 2004).

The purpose of this paper was to analyse the amount of research produced, bibliometric impact, and scientific collaboration in *Online Information Reviews* using bibliometric and social network analyses. The bibliometric description of the papers included in this journal, including the identification of research groups that lead the field in the scope of *Online Information Review,* can help newcomers explore and become familiar with authors, institutions and topics published and provide appropriate understanding that can guide future decisions regarding their research and publication policies.

## 2. Methods

### 2.1. Selection of database and search strategy

We choose Web of Science Core Collection (WoS) to select records from *Online Information Review* and conduct our bibliometric analysis because of the advantages of this database. First, WoS includes copious bibliographical information capable of being reviewed for bibliometric purposes, including all signatory authors of articles and their institutional affiliations, allowing for productivity and social network analysis. Second, WoS details citations received in later papers, which creates an indirect measure of a paper's impact and quality.

Using the search strategy "SO=Online Information Review" in the Advanced Search of WoS and TIMESPAN from 2000 to 2014, we recovered 758 records that were exported to a relational database using the proprietary software *Bibliometricos*.

**2.2. Standardisation of authors and institutional names**

A large amount of information provided in bibliographic databases must be standardised to prevent spelling mistakes, inaccuracies, ambiguities and duplicate records. We were particularly cautious regarding the spelling of names of authors and institutions because some authors and institutions are presented with two or more different names.

**2.3. Bibliometric and social network analysis**

We conducted a bibliometric analysis to identify the annual evolution of published papers and the most productive and most cited authors, institutions and countries, including the ratio of citations per paper. A social network analysis was also conducted to identify groups of collaborating authors, institutions and countries. Publication productivity was considered, giving equal credit to all contributors, e.g., one full unit was assigned to each author, institution or country involved in a paper, as opposed to a fractional count (adjusted count) in which each co-authored paper is divided by the number of co-authors. The software Pajek, designed for the analysis and visualisation of networks, was used for the construction and graphic representation of the research groups (Batagelj and Mrvar, 2001).

To focus the analysis on the more intense collaboration relationships and the legibility and interpretation of graphs, the analysis defined group collaboration as 2 or more authors, institutions or countries jointly signing at least 2 papers (spheres in gold). In figures 3 to 6, the size of the spheres is proportional to the number of papers written in collaboration, and the numeric label accompanying the spheres represents this number. In addition, the thickness of lines connecting two spheres and the numeric label represent the number of papers that were published in collaboration.

# 3. Results

## 3.1. General data: Annual productivity

From 2000 to 2014, 758 articles from *Online Information Review* were included in WoS: 680 original articles (89.7%), 69 editorials (9.1%), 7 reviews and 2 letters (0.3%). The average number of papers per year was 50, rendering 2009 the most productive year (n=67) (Table 1).

*Table 1. Annual evolution of published papers*

| Year | Document Type | | | | |
|---|---|---|---|---|---|
| | Article | Editorial | Letter | Review | Total |
| 2000 | 45 | 10 | 1 | 0 | 56 |
| 2001 | 38 | 2 | 0 | 0 | 40 |
| 2002 | 35 | 6 | 0 | 0 | 41 |
| 2003 | 39 | 3 | 0 | 1 | 43 |
| 2004 | 41 | 5 | 0 | 0 | 46 |
| 2005 | 39 | 4 | 0 | 0 | 43 |
| 2006 | 38 | 6 | 0 | 2 | 46 |
| 2007 | 46 | 7 | 0 | 1 | 54 |
| 2008 | 49 | 6 | 0 | 1 | 56 |
| 2009 | 64 | 3 | 0 | 0 | 67 |
| 2010 | 50 | 1 | 0 | 1 | 52 |
| 2011 | 48 | 4 | 1 | 1 | 54 |
| 2012 | 49 | 4 | 0 | 0 | 53 |
| 2013 | 48 | 5 | 0 | 0 | 53 |
| 2014 | 51 | 3 | 0 | 0 | 54 |
| Total | 680 | 69 | 2 | 7 | 758 |

## 3.2. Authors' productivity and citations

Papers were published by 1097 different authors (Table 2), the most productive being Peter Jacso, from the University of Hawaii Manoa (Hawaii, USA) (n=70); Gorman, from Asia New Zealand Information Associates (Wellington, New Zealand) (n=45); Goh Dion Hoe-Lian from Nanyang Technological University (Nanyang, Singapore) (n=11); Mike Thelwall (Wolverhampton University, Wolverhampton, Midlands, United Kingdom) (n=10); and A.D. Smith (Robert Morris University, Pittsburgh, PA, USA) (n=10). Peter Jacso is also the researcher with the most citations (n=520), followed by Mike Thelwall (n=132). The average number of citations per paper is highest for Flavian (University of Zaragoza) (C/P=14.2), followed by Thelwall (C/P=13.2) and Vaughan (Liwen University, Western Ontario, Canada) (C/P=10). The number of researchers with 0 citations was 330.

*Table 2. Authors with more than 3 published papers and citations*

| Author | Papers | Citations | Average citations / paper |
|---|---|---|---|
| Jacso, Peter | 79 | 520 | 6.58 |
| Gorman, GE | 45 | 55 | 1.22 |
| Goh, Dion Hoe-Lian | 11 | 56 | 5.09 |
| Thelwall, Mike | 10 | 132 | 13.20 |
| Smith, AD | 10 | 59 | 5.90 |
| Bonson-Ponte, Enrique | 8 | 41 | 5.13 |
| Bar-Ilan, Judit | 8 | 40 | 5,00 |
| Tsai, Chih-Fong | 7 | 15 | 2.14 |
| Lee, Kun Chang | 7 | 19 | 2.71 |
| Chua, Alton Yeow-Kuan | 6 | 53 | 8.83 |
| Zhang, Jin | 6 | 19 | 3.17 |
| Lee, Chei Sian | 6 | 24 | 4.00 |
| Flavian, Carlos | 5 | 71 | 14.20 |
| Flores-Munoz, Francisco | 5 | 40 | 8.00 |
| Lu, Hsi-Peng | 5 | 42 | 8.40 |
| Shiri, Ali | 5 | 30 | 6.00 |
| Hui, Siu Cheung | 5 | 13 | 2.60 |
| Zhitomirsky-Geffet, Maayan | 5 | 2 | 0.40 |
| Kochtanek, TR | 5 | 13 | 2.60 |
| Vaughan, Liwen | 4 | 40 | 10.00 |
| Voigt, Kristina | 4 | 24 | 6.00 |
| Na, Jin-Cheon | 4 | 12 | 3.00 |
| Sanz-Blas, Silvia | 4 | 23 | 5.75 |
| Escobar-Rodriguez, Tomas | 4 | 29 | 7.25 |
| Ruiz-Mafe, Carla | 4 | 23 | 5.75 |
| Xie, Hong (Iris) | 4 | 21 | 5.25 |
| Shoham, Snunith | 4 | 19 | 4.75 |
| Shapira, Bracha | 4 | 4 | 1.00 |
| Liew, Chern Li | 4 | 21 | 5.25 |
| Lewandowski, Dirk | 4 | 19 | 4.75 |
| Spink, Amanda | 4 | 31 | 7.75 |
| Ozmutlu, Seda | 4 | 37 | 9.25 |

## 3.3. Institutions' productivity and citations

Of the 453 institutions identified, 12 had published 10 or more articles (Table 3). The majority of the institutions were located in the United States, Spain, Taiwan and the United Kingdom. The University of Hawaii Manoa (Hawaii, USA), ranked highest in institutional productivity with 77 articles, followed by Victoria University of Wellington (New Zealand) (n=34), Nanyang Technological University (Nanyang, Singapore) (n=25) and the University of Zaragoza (Spain) (n=16). These findings highlight the significant number of Asian institutions from Taiwan, South Korea, China and Iran.

*Table 3. Institutions with 5 or more published papers and citations*

| Institutions | Country | Papers | Citations | Average citations / paper |
|---|---|---|---|---|
| University of Hawaii at Manoa | United States of America | 77 | 501 | 6.5 |
| Victoria University of Wellington | New Zealand | 34 | 141 | 4.14 |
| Nanyang Technological University in Singapore | Singapore | 25 | 121 | 4.84 |
| Universidad de Zaragoza | Spain | 16 | 139 | 8.68 |
| Universidad de Granada | Spain | 13 | 40 | 3.07 |
| National Central University in Taiwan | Taiwan | 13 | 47 | 3.61 |
| National Taiwan University | Taiwan | 12 | 23 | 1.91 |
| Bar-Ilan University | Israel | 12 | 26 | 2.16 |
| National Chung Cheng University | Taiwan | 12 | 23 | 1.91 |
| University of Wisconsin–Milwaukee | United States of America | 11 | 43 | 3.9 |
| University of Wolverhampton | United Kingdom | 11 | 146 | 13.27 |
| Robert Morris University in Pittsburgh | United States of America | 10 | 55 | 5.5 |
| University of Western Ontario | Canada | 8 | 46 | 5.75 |
| National Taiwan University of Science and Technology | Taiwan | 8 | 42 | 5.25 |
| Universidad de Huelva | Spain | 8 | 41 | 5.12 |
| University of Strathclyde | United Kingdom | 8 | 75 | 9.37 |
| Wuhan University | Peoples R China | 8 | 7 | 0.87 |
| Korea Advanced Institute of Science and Technology (KAIST) | South Korea | 7 | 11 | 1.57 |
| Islamic Azad University | Iran | 7 | 11 | 1.57 |

| Institutions | Country | Papers | Citations | Average citations / paper |
|---|---|---|---|---|
| Universitat de València | Spain | 7 | 61 | 8.71 |
| GSF National Research Center for Environment and Health, Neuherberg | Germany | 6 | 30 | 5 |
| National Taiwan Normal University | Taiwan | 6 | 17 | 2.83 |
| National Cheng Kung University in Taiwan | Taiwan | 6 | 111 | 18.5 |
| Sungkyunkwan University in Seoul | South Corea | 6 | 25 | 4.16 |
| Universidad de Salamanca | Spain | 5 | 21 | 4.2 |
| Brunel University | United Kingdom | 5 | 27 | 5.4 |
| University of Missouri | United States of America | 5 | 13 | 2.6 |
| University of Ljubljana | Slovenia | 5 | 20 | 4 |
| University of Pittsburgh | United States of America | 5 | 24 | 4.8 |
| National Sun Yat-sen University in Kaohsiung | Taiwan | 5 | 68 | 13.6 |

The University of Hawaii Manoa had the most citations (n=501), followed by the University of Wolverhampton (n=146), Victoria University of Wellington (n=141) and the University of Zaragoza (n=139).

The ratio of citations per article is highest for National Cheng Kung University in Taiwan (C/P=18.5), National Sun Yat-sen University in Kaohsiung (C/P=13.6), the University of Wolverhampton (C/P=13.27) and the University of Strathclyde (United Kingdom) (C/P=9.37). Notably, 120 institutions (26.4%) published papers that were never cited.

**3.4. Countries' productivity and citations**

Of the 54 countries that contributed to the publication of papers (Table 4), the United Sates ranked first with respect to scientific productivity (n=199), followed by Taiwan (n=97), Spain (n=80) and the United Kingdom (n=57). The United Sates also ranked first in citations (n=948), followed by Taiwan (n=481), the United Kingdom (n=403) and Spain (n=393). The highest average citations per paper goes to Denmark (C/P=13.4), Greece (C/P=11.67), Sweden (C/P=11.43) and Finland (C/P=8.17).

*Table 4. Countries with 3 or more published papers and citations*

| Country | Papers | Citations | Average citations / paper |
|---|---|---|---|
| United States of America | 199 | 948 | 4.76 |
| Taiwan | 97 | 481 | 4.96 |
| Spain | 80 | 393 | 4.91 |
| United Kingdom | 57 | 403 | 7.07 |
| New Zealand | 39 | 167 | 4.28 |
| Peoples R China | 34 | 97 | 2.85 |
| South Corea | 27 | 55 | 2.04 |
| Germany | 26 | 144 | 5.54 |
| Singapore | 25 | 121 | 4.84 |
| Canada | 24 | 120 | 5.00 |
| Australia | 21 | 66 | 3.14 |
| Israel | 19 | 52 | 2.74 |
| Iran | 13 | 28 | 2.15 |
| India | 12 | 47 | 3.92 |
| Italy | 7 | 11 | 1.57 |
| Turkey | 7 | 42 | 6.00 |
| Sweden | 7 | 80 | 11.43 |
| South Africa | 6 | 6 | 1.00 |
| Finland | 6 | 49 | 8.17 |
| Slovenia | 6 | 20 | 3.33 |
| Portugal | 5 | 11 | 2.20 |
| Netherlands | 5 | 16 | 3.20 |
| Denmark | 5 | 67 | 13.40 |
| Saudi Arabia | 4 | 13 | 3.25 |
| Greece | 3 | 35 | 11.67 |
| France | 3 | 17 | 5.67 |
| Jordan | 3 | 4 | 1.33 |
| Switzerland | 3 | 11 | 3.67 |
| Norway | 3 | 4 | 1.33 |
| Malaysia | 3 | 12 | 4.00 |

**3.5. Highly cited papers**

The 29 studies receiving more than 20 citations are presented in Table 5. The two most cited articles, "Google Scholar: The pros and the cons" (71 citations) and "Deflated, inflated and phantom citation counts" (60 citations), were published by Jacso (University of Hawaii Manoa, USA) in 2005 and 2006, respectively. Two papers follow with the same number of received citations (n=46), one published in 2008 by Chang and Chen (National Cheng Kung University [NCKU], Taiwan) entitled "The impact of online store environment cues on purchase intention trust and perceived risk as a mediator", and the other paper published in 2001 by Cullen (Victoria University of New Zealand), entitled "Addressing the digital divide". Papers with more than 10 citations are presented in Annex 1.

*Table 5. Highly cited papers*

| Authors | Title | Year | Volume | Number | First Page | Last Page | Web of Science cites |
|---|---|---|---|---|---|---|---|
| Jacso, P | Google Scholar: the pros and the cons | 2005 | 29 | 2 | 208 | 214 | 71 |
| Jacso, P | Deflated, inflated and phantom citation counts | 2006 | 30 | 3 | 297 | 309 | 60 |
| Chang, HH; Chen, SW | The impact of online store environment cues on purchase intention Trust and perceived risk as a mediator | 2008 | 32 | 6 | 818 | 841 | 46 |
| Cullen, R | Addressing the digital divide | 2001 | 25 | 5 | 311 | 320 | 46 |
| Liao, CH; Tsou, CW; Huang, MF | Factors influencing the usage of 3G mobile services in Taiwan | 2007 | 31 | 6 | 759 | 774 | 45 |
| Lee, YC | An empirical investigation into factors influencing the adoption of an e-learning system | 2006 | 30 | 5 | 517 | 541 | 43 |
| Jacso, P | The pros and cons of computing the h-index using Google Scholar | 2008 | 32 | 3 | 437 | 452 | 39 |
| Angus, E; Thelwall, M; Stuart, D | General patterns of tag usage among university groups in Flickr | 2008 | 32 | 1 | 89 | 101 | 36 |
| Jacso, P | Metadata mega mess in Google Scholar | 2010 | 34 | 1 | 175 | 191 | 33 |
| Korfiatis, NT; Poulos, M; Bokos, G | Evaluating authoritative sources using social networks: an insight from Wikipedia | 2006 | 30 | 3 | 252 | 262 | 32 |
| Razmerita, L; Kirchner, K; Sudzina, F | Personal knowledge management The role of Web 2.0 tools for managing knowledge at individual and organisational levels | 2009 | 33 | 6 | 1021 | 1039 | 32 |
| Gavel, Y; Iselid, L | Web of Science and Scopus: a journal title overlap study | 2008 | 32 | 1 | 8 | 21 | 32 |
| Jacso, P | The plausibility of computing the h-index of scholarly productivity and impact using reference-enhanced databases | 2008 | 32 | 2 | 266 | 283 | 32 |
| Casalo, L; Flavian, C; Guinaliu, M | The impact of participation in virtual brand communities on consumer trust and loyalty - The case of free software | 2007 | 31 | 6 | 775 | 792 | 31 |
| Calero, C; Ruiz, J; Piattini, M | Classifying web metrics using the web quality model | 2005 | 29 | 3 | 227 | 248 | 31 |
| Koch, T | Quality-controlled subject gateways: definitions, typologies, empirical overview | 2000 | 24 | 1 | 24 | 34 | 28 |
| Casalo, LV; Flavian, C; Guinaliu, M | The role of security, privacy, usability and reputation in the development of online banking | 2007 | 31 | 5 | 583 | 603 | 28 |
| Gandia, JL | Determinants of web site information by Spanish city councils | 2008 | 32 | 1 | 35 | 57 | 26 |

| Authors | Title | Year | | | | |
|---|---|---|---|---|---|---|
| Thelwall, M; Vaughan, L; Cothey, V; Li, XM; Smith, AG | Which academic subjects have most online impact? A pilot study and a new classification process | 2003 | 27 | 5 | 333 | 343 | 26 |
| Torres, L; Pina, V; Royo, S | E-government and the transformation of public administrations in EU countries - Beyond NPM or just a second wave of reforms? | 2005 | 29 | 5 | 531 | 553 | 25 |
| Goh, DHL; Chua, A; Khoo, DA; Khoo, EBH; Mak, EBT; Ng, MWM | A checklist for evaluating open source digital library software | 2006 | 30 | 4 | 360 | 379 | 25 |
| Jacso, P | The pros and cons of computing the h-index using Web of Science | 2008 | 32 | 5 | 673 | 688 | 24 |
| Jacso, P | The pros and cons of computing the h-index using Scopus | 2008 | 32 | 4 | 524 | 535 | 24 |
| Vasileiou, M; Hartley, R; Rowley, J | An overview of the e-book marketplace | 2009 | 33 | 1 | 173 | 192 | 23 |
| Chen, YN | Application and development of electronic books in an e-Gutenberg age | 2003 | 27 | 1 | 8 | 16 | 22 |
| Cyr, D; Kindra, GS; Dash, S | Web site design, trust, satisfaction and e-loyalty: the Indian experience | 2008 | 32 | 6 | 773 | 790 | 22 |
| Mayr, P; Walter, AK | An exploratory study of Google Scholar | 2007 | 31 | 6 | 814 | 830 | 22 |
| Chiu, CM; Chang, CC; Cheng, HL; Fang, YH | Determinants of customer repurchase intention in online shopping | 2009 | 33 | 4 | 761 | 784 | 22 |
| Rowley, J | Online branding | 2004 | 28 | 2 | 131 | 138 | 22 |

### 3.6. Scientific collaboration patterns

#### 3.6.1. Collaboration Index

Of the 758 analysed papers, 334 (44%) were signed by a single author, 192 (25.3%) by two authors and 159 (21%) by three authors. The collaboration index increased from 1.59 authors per paper in the early 2000s to 2.74 in 2014 (Figure 1).

*Figure 1. Annual evolution of the collaboration Index*

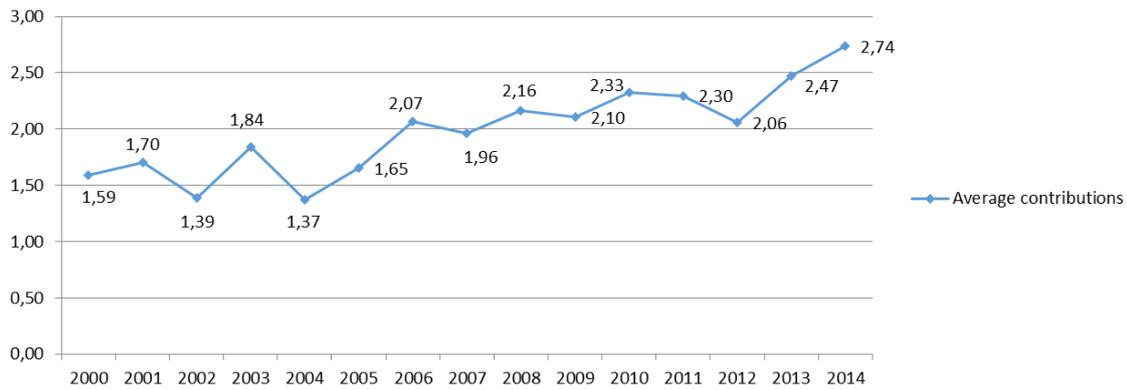

## 3.6.2. Institutional collaboration

Figure 2 shows the evolution of domestic compared with international collaboration as well as papers written without collaboration.

*Figure 2. Institutional collaboration*

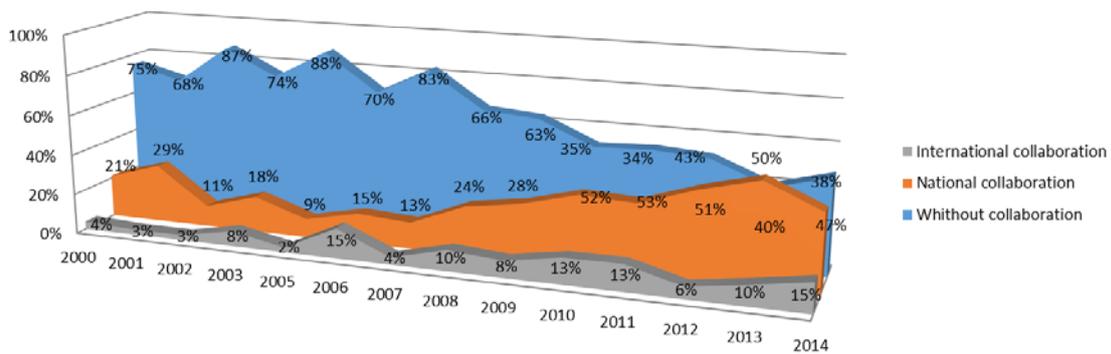

## 3.6.3. Network of collaboration between authors

Figures 3 and 4 show the primary groups integrating the network of collaboration among authors. The first group (Figure 3) is integrated by 10 researchers. The size of the spheres is proportional to the number of published papers by each author, and the size of the lines connecting two countries is proportional to the number of papers published in collaboration. The group integrates 8 researchers from Nanyang Technological University (Singapore Singapore), Goh being the author with more centrality. The second group includes researchers from Israel, and the central author is Shapira (Ben Gurion University Negev). The third group, with 7 researchers, comprises investigators from the University of Valencia, Spain. Two other groups with 7 researchers each involve researchers from the United Kingdom (Norris and Openheim from University Loughborough, Leicestershire, England) and South Korea (Lee, Hanyang University, and Lim, Sungkyunkwan University). The three remaining groups of 6 authors each represent several universities from Taiwan (two groups) and Israel (Bar Ilan University). Figure 4 presents 4 additional groups with 5 components and 7 groups with 4 components. These groups represent Taiwan (two groups from National Taiwan Normal University, Chung Hua University, Hwa Hsia Institute of Technology, National Central University, and National Chung Cheng University), Singapore (Nanyang Technological University), Spain (five groups from universities in Zaragoza, Castilla La Mancha, Granada and Huelva), the United Kingdom (one group from the University of Strathclyde, Glasgow, Lanark, Scotland) and Turkey (one group from Uludag University).

*Figure 3. Network of collaboration between authors. Main groups*

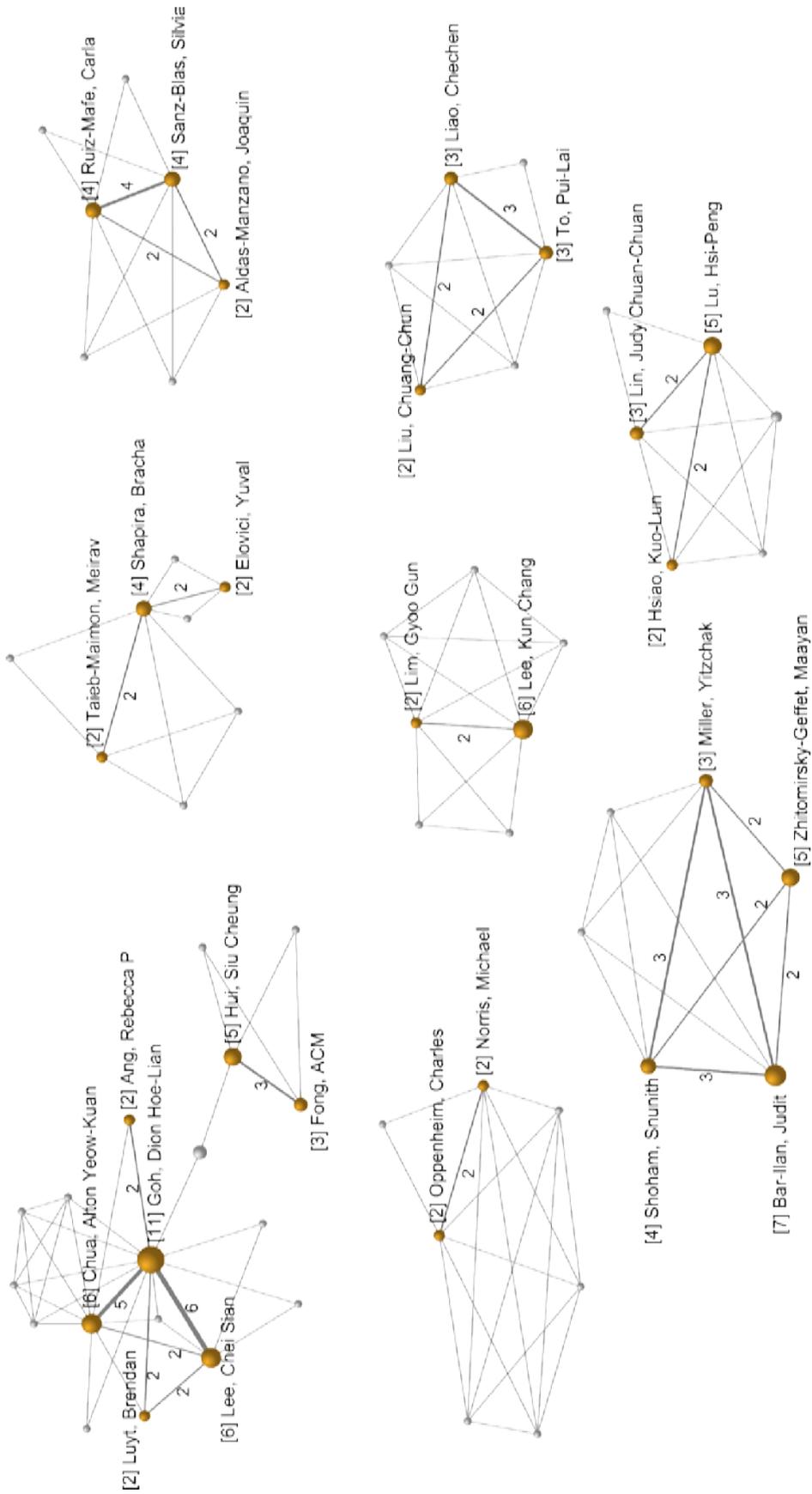

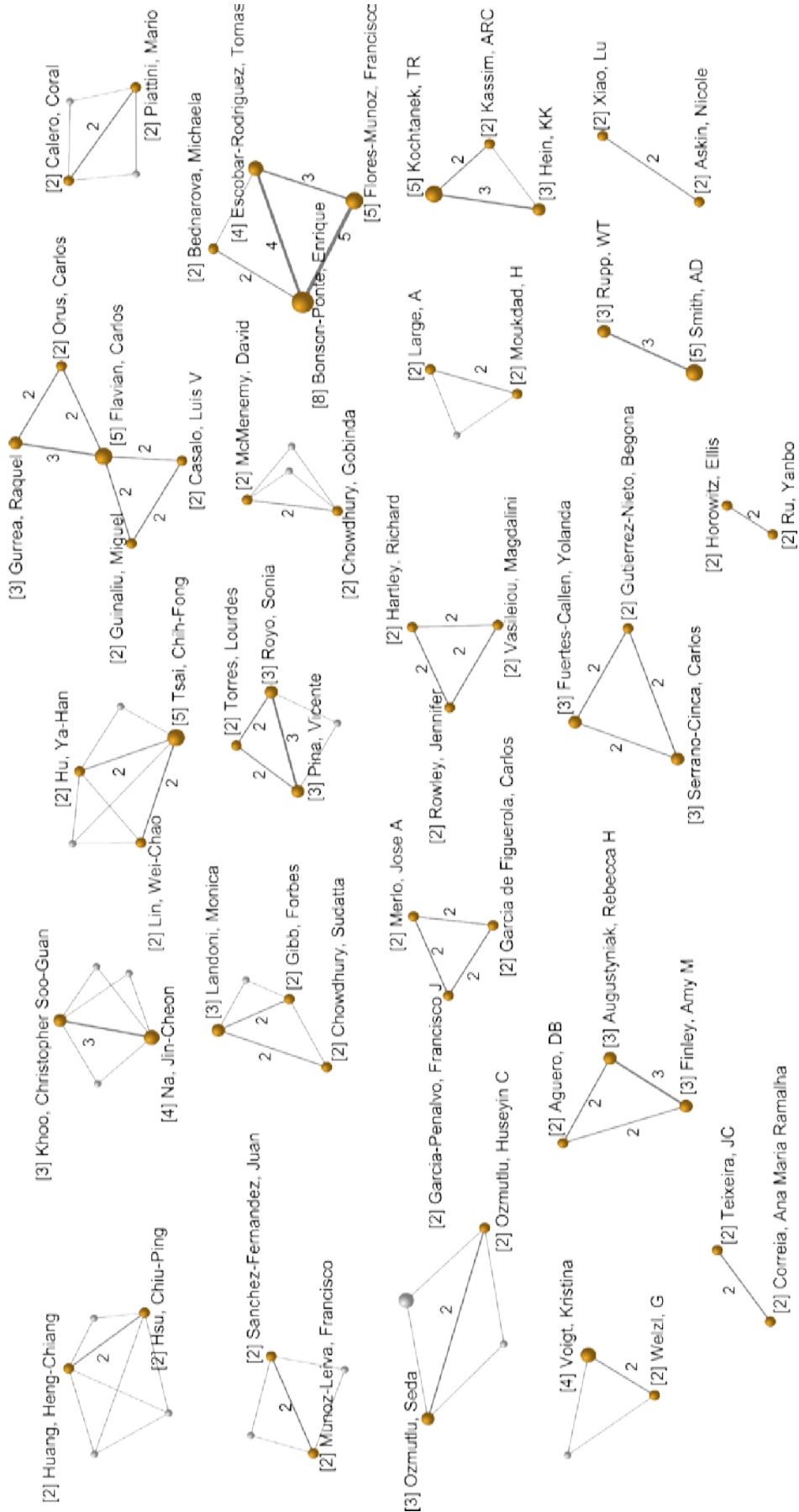

*Figure 4. Network of collaboration between authors. Other groups*

**3.6.4. Network of collaboration among institutions**

The network of collaboration among institutions (Figure 5) shows several non-connected large groups of institutions. The group with the most institutions (n=15) comprises institutions from Taiwan; National Central University, National Chung Cheng University and National Taiwan University are the most central and connect most often with peripheral institutions. The second group integrates institutions from South Korea, with the Korean Advanced Institute of Science and Technology in the most central position. A group of 5 components includes institutions from Portugal. Other groups with 3 components involve institutions from the United Kingdom (Brunel University) and Norway (the Norwegian School of Information Technology). Finally, the network shows three non-connected groups with only two components, two from the United States (one connecting the University of Missouri with the University of Nebraska Omaha and the other group connecting Robert Morris University and the University of Montevallo) and one from Canada (connecting the universities of McGill and Dalhousie).

*Figure 5. Network of collaboration between institutions*

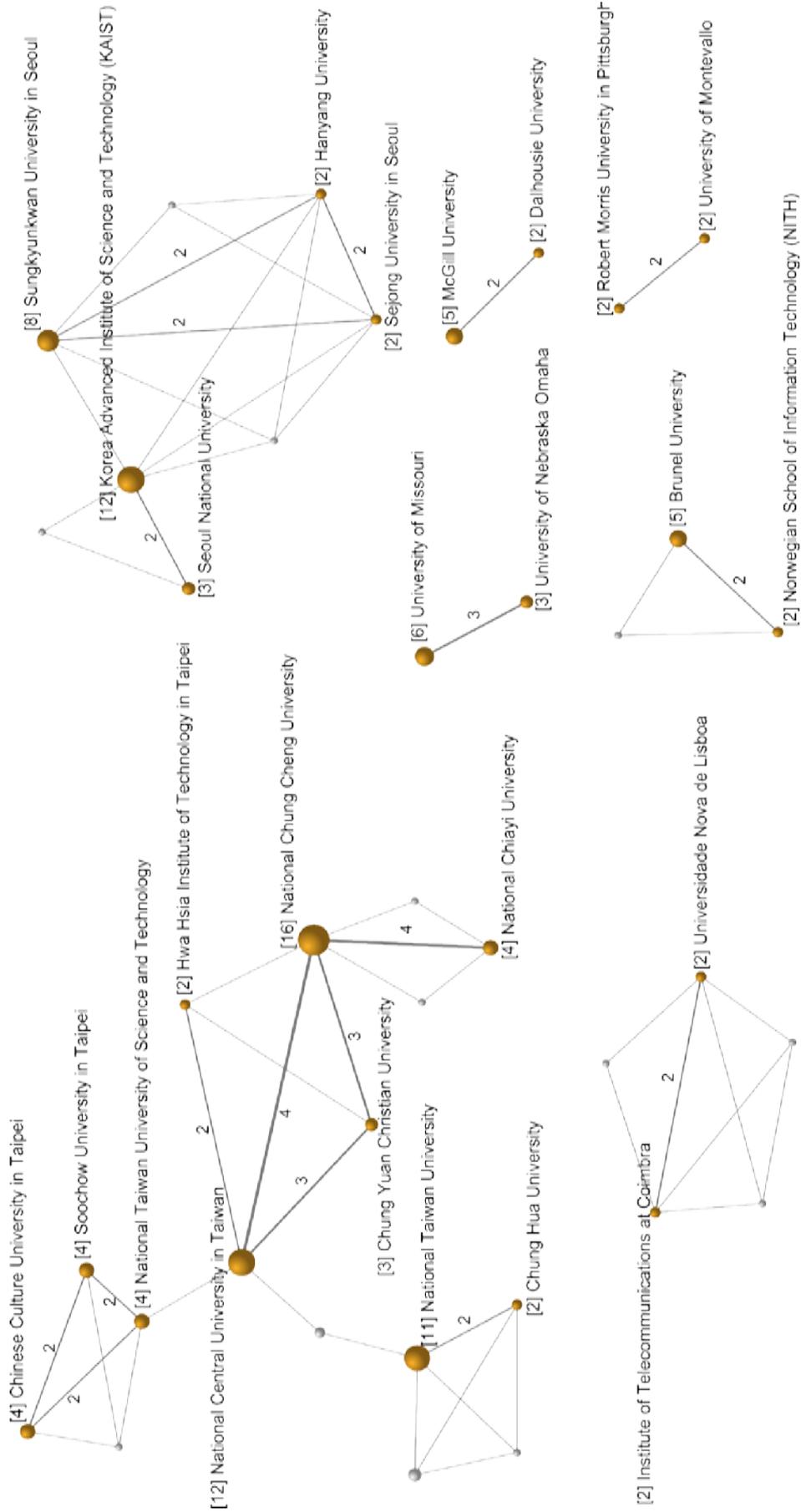

## 3.6.5. Network of collaboration among countries

Figure 6 shows the network of collaboration among countries. The United States is the country with the more central position because the U.S. collaborates with numerous other countries (n=11), followed by the United Kingdom, China, Canada and Spain. The most collaboration has been between the United States and South Korea (n=6).

*Figure 6. Network of collaboration between countries*

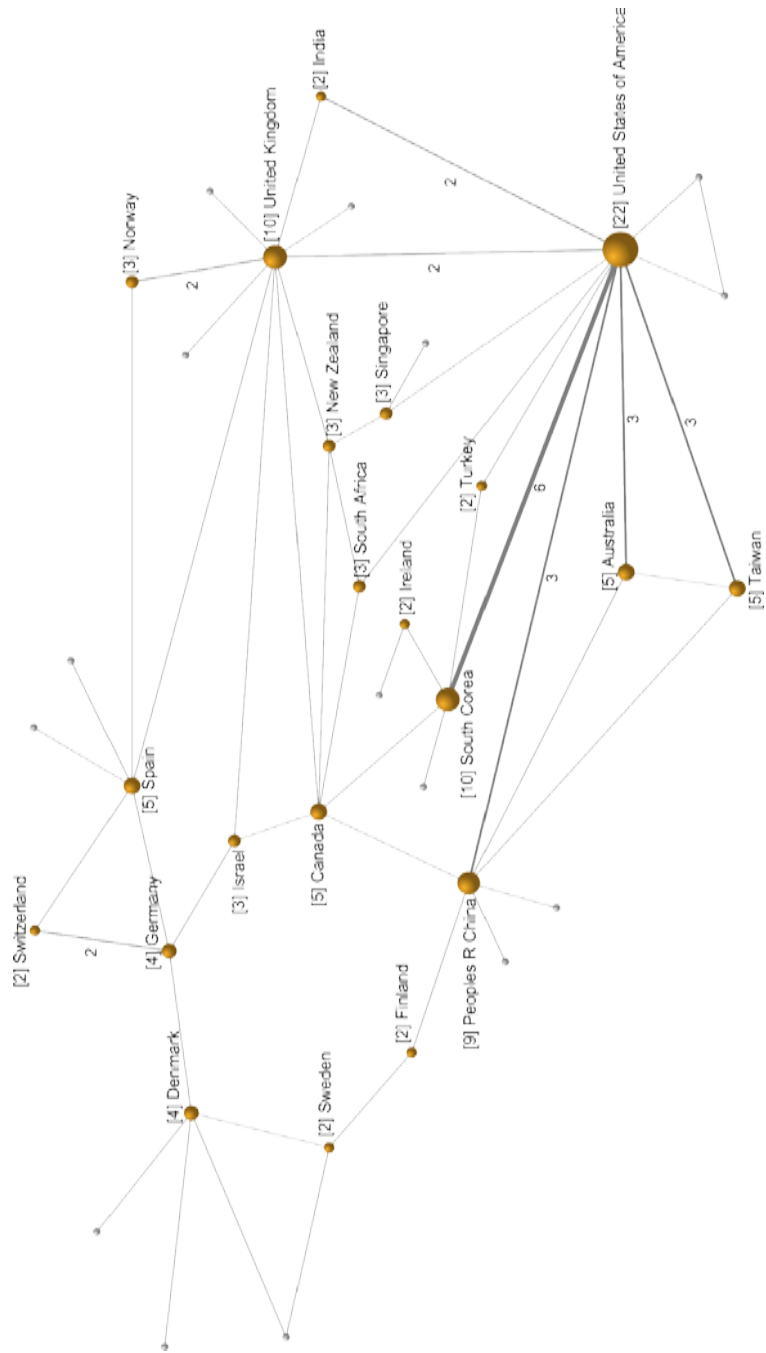

# 4. Discussion

This paper has identified the primary authors, institutions and countries leading the global research on online information published in the *Online Information Review* journal in addition to research groups, institutions and corresponding countries working together in collaboration. Studying a single journal bibliometrically creates a detailed, multi-faceted picture of the characteristics of the journal, providing a description that offers insight that extends beyond the superficial (Nebelong-Bonnevie and Frandsen, 2006). Such analysis provides information on the productivity, visibility and even maturity of a journal in a research field and informs us about the research activity in the field and the research orientation that the journal supports to spread the knowledge for which the journal was created (Anyi *et al*., 2009).

More bibliometric studies on single journals or comparing two or more journals have been published in the field of library and information science. In a review of single journal studies in this area, Tiew (1997) identified 21 studies reported in the literature with 11 unique journal titles because some journals were studied by more than one article. The *Journal of the Association for Information Science and Technology* (JASIST) and the *Journal of Documentation* and *Scientometrics* were analysed several times, reflecting their continued influence and importance in sustaining the interests of bibliometrists over the years. Remarkably, researchers agree that to provide a complete and integrated portrait of a single journal requires combining a variety of bibliometric measures. Those metrics can be used not only to understand the journal's characteristics but also the characteristics of the literature and communication behavior in the fields the journals represent. According to a study by Anyi *et al*. (2009), the primary bibliometric measures used to study single journals have been article productivity, authors' characteristics, authors' productivity, co-authorship patterns, content analysis, citation analysis, and editorial board characteristics.

*Online Information Review* publishes an average of 50 articles annually. That nearly constant number can be considered normal because journals editors do not generally increase the number of annual published papers. Increasing the number of papers can produce unwanted decreases in effect by increasing the denominator of the calculation formula.

As reported in other studies (Koehler *et al.*, 2000), the majority of papers were contributed by professionals affiliated with universities and research centers. Such a trend indicates a shift in recent years from corporate and service-oriented institutions towards university faculties. Notably, in addition to the traditional institutions of Western countries such as the United States, the United Kingdom, Spain and Canada, numerous institutions from Asian countries such as Singapore, Taiwan and China are represented. These contributions are consistent with the emergent development of science in newly industrialized countries.

An important feature to consider is the degree of internationalization of authors who publish in a journal, and most journals welcome the chance to have a colourful mixture of authors from all over the world. A reasonable variety of authors from various institutions from many countries reveals solid editorial policy and may also encourage authors to choose that journal in which to publish. As shown in our work, *Online Information Review* is a journal with a great variety of participating institutions from countries around the world, representing 453 institutions from 50 different countries.

The analysis of the content of highly cited papers provided useful information on the topics that generate interest among researchers. Such topics include the pros and cons of Google Scholar, citation counts and h-index, online and digital store environments, 3G mobile services, e-learning systems, web 2.0, and metrics based on the web.

Collaborative research can be effectively measured by the number of authors signing a set of papers. The percentage of single-authored papers in *Online Information Review* (44%) is lower than the percentage identified by Al-Ghamdi *et al.* (1998) in JASIST from 1970 to 1996 (61%) and by Naqvi (2005) in *Journal of Documentation* between 1994 and 2003 (55.8%). However, this percentage was shown to be lower than Liu's findings in 2003 and nearer to the *Online Information Review*'s reporting (42.3%). The collaboration index of *Online Information Review* has increased from 1.59 authors per paper in the early 2000s to 2.74 in 2014. These data may indicate that this field, which was not highly collaborative in past decades, is evolving towards greater cooperation.

This increase in papers by multiple authors has also been reported by Lipetz (1999), Nisonger (1999) and Liu (2003) in other studies based on JASIST. The emergence of multi-authored works may indicate the shift observed by Koehler *et al.* (2000) from articles derived from non-funded research by single authors to articles that are increasingly funded and multi-authored from various regions or countries. Those researchers also suggested that this trend reflects a more complex and cross-fertilized research activity and a healthy feature of information science as a discipline in itself, i.e., the convergence of scientific disciplines and the interconnected teamwork of many researchers from many institutions or fields towards the same goal (Musek *et al.*, 2003).

The collaboration between countries shows some patterns different from patterns observed in other journals. In addition to the traditional collaboration between the US and the United Kingdom in most research fields, the *Online Information Review* also contains the most collaboration between the US and Asian countries such as South Korea, India, Taiwan and Singapore. The United Kingdom serves as the connection between the US and important producing countries such as Spain, Canada and Germany. The output of China and the collaboration between China and the US are also quite significant. Notably, studies by He (2001) and He and Spink (2002) revealed that authors from the United Kingdom and Canada contributed the most to JASIST and authors from the United States and Canada contributed the most to the Journal of Documentation.

Collaboration among authors, institutions and countries has been facilitated by more cost-effective and efficient international travel and communication, the widespread use of information media, a larger nucleus of more highly qualified researchers, and competition for employment and funding opportunities because many funding agencies are giving precedence to collaboration when allocating grants (Wiles *et al.*, 2011).

## 4.1. Limitations

This study did not account for every article published in every issue of the journal. Therefore, some bibliometric data will have been missed. However, the samplings analysed are the most representative papers in the journal and have allowed us to draw conclusions regarding actors and trends across time. Citation counts were conducted in Web of Science in 2014, and it is possible that the number of citations has currently increased.

**4.2. Future Work**

Replicating this work in a few years to observe changes in the publication patterns of this journal would be interesting. An evident comparison with other journals with influence in the same research area may also be unavoidable.

# 5. Conclusions

The findings of the present study on *Online Information Review* should be of great interest to library and information science professionals and a great help in understanding the nature of research information science. In addition, it would be helpful for the journal editor to obtain the bibliometric portrait of the studied journal and recognize the interactions between authors, institutions and countries. Assessing the bibliometric profile of a journal (as a depository of research) may allow identification of the journal's publication practices. The identification of research groups that publish in *Online Information Review* can help beginners explore the field and become familiar with groups of authors and institutions and can provide appropriate understanding to guide future decisions regarding their research and publication policies. A move towards greater collaboration among authors has been observed. This study should be a 'bibliometric barometer' of the current state of research in the area.

# 6. References

Aleixandre-Benavent, R., Valderrama Zurian, J.C. and González Alcaide, G. (2007), "The impact factor of scientific journals: Limitations and alternative indicators", *El Profesional de la Información*, Vol. 16, pp. 4-11.

Al-Ghamdi, A., Al-Harbi, M., Beacom, N.A.B., Dedolph, J., Deignan, M., Elftmann, C., Finley, N., LoCicero, L., Middlecamp, J., O'Regan, C., Pluskota, F., Ritter, A.A., Russell, S., Sabat, I., Schneider, J., Schoeberl, M., Tragash, P. and Withers, B.H.

(1998), "Authorship in JASIS: A quantitative analysis", *Katherine Sharp Review*, Vol. 6, available at: https://web.archive.org/web/20040704111510/http://www.lis.uiuc.edu/review/6/al_ghamdi.pdf (accessed 29 July 2016).

Anyi, K.W.U., Zainab, A.N. and Anuar, N.B. (2009), "Bibliometric studies on single journals: a review", *Malaysian Journal of Library & Information Science*, Vol. 14, No. 1, pp. 17-55.

Batagelj, V. and Mrvar, A. (2001), *Pajek Program for Large Network Analysis*, University of Ljubljana, Slovenia.

González-Alcaide, G., Valderrama-Zurián, J.C. and Aleixandre-Benavent, R. (2008a), "Research fronts and collaboration patterns in Reproductive Biology. Coauthorship networks and institutional collaboration", *Fertility and Sterility*, Vol. 90, pp. 941-956.

González-Alcaide, G., Alonso Arroyo, A., González De Dios, J., Pérez Sempere, A.P., Valderrama-Zurián, J.C. and Aleixandre-Benavent, R. (2008b), "Co-authorship networks and institutional collaboration in Revista de Neurología", *Revista de Neurologia*, Vol. 46, pp. 642-651.

He, S. (2001), "Geographic distribution of foreign authorship in LIS journals: A comparison between JASIST and Journal of Documentation", in Davis, M. and Wilson, C. (Eds.), *Proceedings of the 8th international conference on scientometrics & informetrics*, Bibliometric & Informetric Research Group, Sydney, pp. 235-243.

He, S. and Spinks, A. (2002), "A comparison of foreign authorship distribution in JASIST and the Journal of Documentation", *Journal of the American Society of Information Science and Technologist*, Vol. 53, No. 11, pp. 953-959.

Koehler, W., Anderson, A.D., Beverly, A.D., Fields, D.E., Golden, M.H., Dawn, J., Johnson, A.C., Kipp, C., Ortega, L.L., Ripley, E.B., Roddy, R.L., Shaffer, K.B., Shelburn, S. and Wasteneys, C.D. (2000), "A profile in statistics of journal articles:

Fifty years of American Documentation and the Journal of the American Society for Information Science", *Cybermetrics*, Vol. 4, No. 1, Paper 3.

Kretschmer, H. (2004), "Author productivity and geodesic distance in bibliographic co-authorship networks, and visibility on the Web", *Scientometrics*, Vol. 60, No. 3, pp. 409-420.

Lipetz, B.A. (1999), "Aspects of JASIS authorship through five decades", *Journal of the American Society for Information Science*, Vol. 50, No. 11, pp. 994-1003.

Liu, J. (2003), "A bibliometric study: author productivity and co-authorship features of JASIST 2001-2002", *Mississippi Libraries*, Vol. 67, No. 4, pp. 110-112.

Musek, M., Oven, M. and Juinic, P. (2003), "Ten years of the journal Radiology and Oncology some bibliometric evaluations", *Radiology and Oncology*, Vol. 37, No. 3, pp. 141-153.

Naqvi, S.H. (2005), "Journal of Documentation: a bibliometric study", *International Information Communication and Education*, Vol. 24, No. 1, pp. 53-56.

Nebelong-Bonnevie, E. and Frandsen, T.F. (2006), "Journal citation identity and journal citation image: a portrait of the Journal of Documentation", *Journal of Documentation*, Vol. 62, No. 1, pp. 30-57.

Newman, M.E.J. (2004), "Coauthorship networks and patterns of scientific collaboration", *Proceedings of the National Academy of Science of the United States of America*, Vol. 101, pp. 5200-5205.

Nisonger, T.E. (1999), "JASIS and library and information science journal rankings: a review and analysis of the last half-century", *Journal of the American Society for Information Science*, Vol. 50, No. 11, pp. 1004-1019.

Tiew, W.S. (1997), "Single journal bibliometric studies: a review", *Malaysian Journal of Library & Information Science*, Vol. 2, No. 2, pp. 93-114.

White, H.D. and McCain, K.W. (1989), "Bibliometrics", *Annual Review of Information Science and Technology*, Vol. 24, pp. 119-186.

Wiles, L., Olds, T. and Williams, M. (2010), "Twenty-five years of Australian nursing and allied health professional journals: bibliometric analysis from 1985 through 2010", *Scientometrics*, Vol. 94, No. 1, pp. 359-378.

# 7. Appendix

### Annex 1. Most cited papers (n>10 citations)

| Author | Title | Year | Volume | Number | First Page | Last Page | Web of Science cites |
|---|---|---|---|---|---|---|---|
| Jacso, P | Google Scholar: the pros and the cons | 2005 | 29 | 2 | 208 | 214 | 71 |
| Jacso, P | Deflated, inflated and phantom citation counts | 2006 | 30 | 3 | 297 | 309 | 60 |
| Chang, HH; Chen, SW | The impact of online store environment cues on purchase intention Trust and perceived risk as a mediator | 2008 | 32 | 6 | 818 | 841 | 46 |
| Cullen, R | Addressing the digital divide | 2001 | 25 | 5 | 311 | 320 | 46 |
| Liao, CH; Tsou, CW; Huang, MF | Factors influencing the usage of 3G mobile services in Taiwan | 2007 | 31 | 6 | 759 | 774 | 45 |
| Lee, YC | An empirical investigation into factors influencing the adoption of an e-learning system | 2006 | 30 | 5 | 517 | 541 | 43 |
| Jacso, P | The pros and cons of computing the h-index using Google Scholar | 2008 | 32 | 3 | 437 | 452 | 39 |
| Angus, E; Thelwall, M; Stuart, D | General patterns of tag usage among university groups in Flickr | 2008 | 32 | 1 | 89 | 101 | 36 |
| Jacso, P | Metadata mega mess in Google Scholar | 2010 | 34 | 1 | 175 | 191 | 33 |
| Korfiatis, NT; Poulos, M; Bokos, G | Evaluating authoritative sources using social networks: an insight from Wikipedia | 2006 | 30 | 3 | 252 | 262 | 32 |
| Razmerita, L; Kirchner, K; Sudzina, F | Personal knowledge management The role of Web 2.0 tools for managing knowledge at individual and organisational levels | 2009 | 33 | 6 | 1021 | 1039 | 32 |
| Gavel, Y; Iselid, L | Web of Science and Scopus: a journal title overlap study | 2008 | 32 | 1 | 8 | 21 | 32 |
| Jacso, P | The plausibility of computing the h-index of scholarly productivity and impact using reference-enhanced databases | 2008 | 32 | 2 | 266 | 283 | 32 |
| Casalo, L; Flavian, C; Guinaliu, M | The impact of participation in virtual brand communities on consumer trust and loyalty - The case of free software | 2007 | 31 | 6 | 775 | 792 | 31 |
| Calero, C; Ruiz, J; Piattini, M | Classifying web metrics using the web quality model | 2005 | 29 | 3 | 227 | 248 | 31 |

| Author | Title | Year | Volume | Number | First Page | Last Page | Web of Science cites |
|---|---|---|---|---|---|---|---|
| Koch, T | Quality-controlled subject gateways: definitions, typologies, empirical overview | 2000 | 24 | 1 | 24 | 34 | 28 |
| Casalo, LV; Flavian, C; Guinaliu, M | The role of security, privacy, usability and reputation in the development of online banking | 2007 | 31 | 5 | 583 | 603 | 28 |
| Gandia, JL | Determinants of web site information by Spanish city councils | 2008 | 32 | 1 | 35 | 57 | 26 |
| Thelwall, M; Vaughan, L; Cothey, V; Li, XM; Smith, AG | Which academic subjects have most online impact? A pilot study and a new classification process | 2003 | 27 | 5 | 333 | 343 | 26 |
| Torres, L; Pina, V; Royo, S | E-government and the transformation of public administrations in EU countries - Beyond NPM or just a second wave of reforms? | 2005 | 29 | 5 | 531 | 553 | 25 |
| Goh, DHL; Chua, A; Khoo, DA; Khoo, EBH; Mak, EBT; Ng, MWM | A checklist for evaluating open source digital library software | 2006 | 30 | 4 | 360 | 379 | 25 |
| Jacso, P | The pros and cons of computing the h-index using Web of Science | 2008 | 32 | 5 | 673 | 688 | 24 |
| Jacso, P | The pros and cons of computing the h-index using Scopus | 2008 | 32 | 4 | 524 | 535 | 24 |
| Vasileiou, M; Hartley, R; Rowley, J | An overview of the e-book marketplace | 2009 | 33 | 1 | 173 | 192 | 23 |
| Chen, YN | Application and development of electronic books in an e-Gutenberg age | 2003 | 27 | 1 | 8 | 16 | 22 |
| Cyr, D; Kindra, GS; Dash, S | Web site design, trust, satisfaction and e-loyalty: the Indian experience | 2008 | 32 | 6 | 773 | 790 | 22 |
| Mayr, P; Walter, AK | An exploratory study of Google Scholar | 2007 | 31 | 6 | 814 | 830 | 22 |
| Chiu, CM; Chang, CC; Cheng, HL; Fang, YH | Determinants of customer repurchase intention in online shopping | 2009 | 33 | 4 | 761 | 784 | 22 |
| Rowley, J | Online branding | 2004 | 28 | 2 | 131 | 138 | 22 |
| Jacso, P | Errors of omission and their implications for computing scientometric measures in evaluating the publishing productivity and impact of countries | 2009 | 33 | 2 | 376 | 385 | 20 |
| Salo, J; Karjaluoto, H | A conceptual model of trust in the online environment | 2007 | 31 | 5 | 604 | 621 | 20 |
| Chu, SKW | TWiki for knowledge building and management | 2008 | 32 | 6 | 745 | 758 | 19 |
| Bonson-Ponte, E; Escobar-Rodriguez, T; Flores-Munoz, F | Online transparency of the banking sector | 2006 | 30 | 6 | 714 | 730 | 18 |
| Bar-Ilan, J | Evaluating the stability of the search tools Hotbot and Snap: a case study | 2000 | 24 | 6 | 439 | 449 | 18 |

| Author | Title | Year | Volume | Number | First Page | Last Page | Web of Science cites |
|---|---|---|---|---|---|---|---|
| Kani-Zabihi, E; Ghinea, G; Chen, SY | Digital libraries: what do users want? | 2006 | 30 | 4 | 395 | 412 | 18 |
| Lee, MC | Understanding the behavioural intention to play online games An extension of the theory of planned behaviour | 2009 | 33 | 5 | 849 | 872 | 17 |
| Gilchrist, A | Corporate taxonomies: report on a survey of current practice | 2001 | 25 | 2 | 94 | 102 | 17 |
| Stoddart, L | Managing intranets to encourage knowledge sharing: opportunities and constraints | 2001 | 25 | 1 | 19 | 28 | 17 |
| Thelwall, M | Research dissemination and invocation on the Web | 2002 | 26 | 6 | 413 | 420 | 17 |
| Lu, HP; Hsiao, KL | Gender differences in reasons for frequent blog posting | 2009 | 33 | 1 | 135 | 156 | 16 |
| Thelwall, M | Blog searching - The first general-purpose source of retrospective public opinion in the social? | 2007 | 31 | 3 | 277 | 289 | 16 |
| Yeh, YS; Li, YM | Building trust in m-commerce: contributions from quality and satisfaction | 2009 | 33 | 6 | 1066 | 1086 | 16 |
| Jacso, P | Five-year impact factor data in the Journal Citation Reports | 2009 | 33 | 3 | 603 | 614 | 16 |
| Liew, CL; Foo, S; Chennupati, KR | A study of graduate student end-users´ use and perception of electronic journals | 2000 | 24 | 4 | 302 | 315 | 16 |
| Shin, DH | Analysis of online social networks: a cross-national study | 2010 | 34 | 3 | 473 | 495 | 15 |
| Ozmutlu, S; Ozmutlu, HC; Spink, A | Are people asking questions of general Web search engines? | 2003 | 27 | 6 | 396 | 406 | 15 |
| Moukdad, H; Large, A | Users´ perceptions of the Web as revealed by transaction log analysis | 2001 | 25 | 6 | 349 | 358 | 15 |
| Jacso, P | Dubious hit counts and cuckoo´s eggs | 2006 | 30 | 2 | 188 | 193 | 14 |
| Domingo-Ferrer, J; Solanas, A; Castella-Roca, J | h(k)-private information retrieval from privacy-uncooperative queryable databases | 2009 | 33 | 4 | 720 | + | 14 |
| Ru, YB; Horowitz, E | Indexing the invisible web: a survey | 2005 | 29 | 3 | 249 | 265 | 14 |
| Bigne-Alcaniz, E; Ruiz-Mafe, C; Aldas-Manzano, J; Sanz-Blas, S | Influence of online shopping information dependency and innovativeness on internet shopping adoption | 2008 | 32 | 5 | 648 | 667 | 14 |
| Gorman, GE | "They can´t read, but they sure can count" Flawed rules of the journal rankings game | 2008 | 32 | 6 | 705 | 708 | 14 |
| Tillotson, J | Web site evaluation: a survey of undergraduates | 2002 | 26 | 6 | 392 | 403 | 14 |
| Chowdhury, G; Poulter, A; McMenemy, D | Public library 2.0 - Towards a new mission for public libraries as a "network of community knowledge" | 2006 | 30 | 4 | 454 | 460 | 14 |
| Li, XM | A review of the development and application of the Web impact factor | 2003 | 27 | 6 | 407 | 417 | 14 |
| Wang, SM; Lin, JCC | The effect of social influence on bloggers´ usage intention | 2011 | 35 | 1 | 50 | 65 | 14 |

| Author | Title | Year | Volume | Number | First Page | Last Page | Web of Science cites |
|---|---|---|---|---|---|---|---|
| Allen, M | A case study of the usability testing of the University of South Florida´s virtual library interface design | 2002 | 26 | 1 | 40 | 53 | 14 |
| Chowdhury, S; Landoni, M; Gibb, F | Usability and impact of digital libraries: a review | 2006 | 30 | 6 | 656 | 680 | 14 |
| Shiri, AA; Revie, C | Thesauri on the Web: current developments and trends | 2000 | 24 | 4 | 273 | 279 | 13 |
| Su, XN; Han, XM; Han, XN | Developing the Chinese social science citation index | 2001 | 25 | 6 | 365 | 369 | 13 |
| Majid, S; Tan, AT | Usage of information resources by computer engineering students: a case study of Nanyang Technological University, Singapore | 2002 | 26 | 5 | 318 | 325 | 13 |
| Goh, DHL; Chua, A; Lee, CS; Razikin, K | Resource discovery through social tagging: a classification and content analytic approach | 2009 | 33 | 3 | 568 | 583 | 13 |
| Black, EW | Wikipedia and academic peer review - Wikipedia as a recognised medium for scholarly publication? | 2008 | 32 | 1 | 73 | 88 | 13 |
| Ozmutlu, S; Cavdur, F | Neural network applications for automatic new topic identification | 2005 | 29 | 1 | 34 | 53 | 13 |
| Xie, H; Cool, C | Ease of use versus user control: an evaluation of Web and non-Web interfaces of online data bases | 2000 | 24 | 2 | 102 | 115 | 13 |
| Chang, HH; Wang, HW | The moderating effect of customer perceived value on online shopping behaviour | 2011 | 35 | 3 | 333 | 359 | 13 |
| Jacso, P | The h-index for countries in Web of Science and Scopus | 2009 | 33 | 4 | 831 | 837 | 13 |
| Still, JM | A content analysis of university library Web sites in English speaking countries | 2001 | 25 | 3 | 160 | 164 | 12 |
| Chen, XT | MetaLib, WebFeat, and Google - The strengths and weaknesses of federated search engines compared with Google | 2006 | 30 | 4 | 413 | 427 | 12 |
| Zhou, T | Examining the critical success factors of mobile website adoption | 2011 | 35 | 4 | 636 | 652 | 12 |
| Bilal, D | Perspectives on children´s navigation of the World Wide Web: does the type of search task make a difference? | 2002 | 26 | 2 | 108 | 117 | 12 |
| Levine-Clark, M; Gil, E | A comparative analysis of social sciences citation tools | 2009 | 33 | 5 | 986 | 996 | 12 |
| Bar-Ilan, J; Shoham, S; Idan, A; Miller, Y; Shachak, A | Structured versus unstructured tagging: a case study | 2008 | 32 | 5 | 635 | 647 | 12 |
| Jacso, P | Comparison of journal impact rankings in the SCImago Journal & Country Rank and the Journal Citation Reports databases | 2010 | 34 | 4 | 642 | 657 | 12 |
| Norris, M; Oppenheim, C; Rowland, F | Finding open access articles using Google, Google Scholar, OAIster and OpenDOAR | 2008 | 32 | 6 | 709 | 715 | 12 |

| Author | Title | Year | Volume | Number | First Page | Last Page | Web of Science cites |
|---|---|---|---|---|---|---|---|
| Jacso, P | Calculating the h-index and other bibliometric and scientometric indicators from Google Scholar with the Publish or Perish software | 2009 | 33 | 6 | 1189 | 1200 | 12 |
| Smith, AD | Cybercriminal impacts on online business and consumer confidence | 2004 | 28 | 3 | 224 | 234 | 12 |
| Chiu, CM; Wang, ETG; Shih, FJ; Fan, YW | Understanding knowledge sharing in virtual communities An integration of expectancy disconfirmation and justice theories | 2011 | 35 | 1 | 134 | 153 | 11 |
| Voigt, K; Welzl, G | Chemical databases: an overview of selected databases and evaluation methods | 2002 | 26 | 3 | 172 | 192 | 11 |
| Gorman, GE | A tale of information ethics and encyclopaedias; or, is Wikipedia just another internet scam? | 2007 | 31 | 3 | 273 | 276 | 11 |
| Lu, HP; Lee, MR | Demographic differences and the antecedents of blog stickiness | 2010 | 34 | 1 | 21 | 38 | 11 |
| Witten, IH; Bainbridge, D; Boddie, SJ | Greenstone: open-source digital library software with end-user collection building | 2001 | 25 | 5 | 288 | 298 | 11 |
| Huvila, I; Holmberg, K; Ek, S; Widen-Wulff, G | Social capital in Second Life | 2010 | 34 | 2 | 295 | 316 | 11 |
| Bonson, E; Flores, F | Social media and corporate dialogue: the response of global financial institutions | 2011 | 35 | 1 | 34 | 49 | 11 |
| Smith, AD; Manna, DR | Exploring the trust factor in e-medicine | 2004 | 28 | 5 | 346 | 355 | 11 |
| Hsiao, KL; Lin, JCC; Wang, XY; Lu, HP; Yu, HJ | Antecedents and consequences of trust in online product recommendations An empirical study in social shopping | 2010 | 34 | 6 | 935 | 953 | 11 |
| Trkman, M; Trkman, P | A wiki as intranet: a critical analysis using the Delone and McLean model | 2009 | 33 | 6 | 1087 | 1102 | 11 |
| Jacso, P | Database source coverage: hypes, vital signs and reality checks | 2009 | 33 | 5 | 997 | 1007 | 11 |